\documentclass[12pt,a4paper,fleqn]{article}
\usepackage{umlaut,graphicx,epsfig,epsf,rotating}
\newcommand{\etal}{{\sl et. al.~}}
\newcommand{\beq}{\begin{equation}}
\newcommand{\eeq}{\end{equation}}
\newcommand{\HM} {HEIDELBERG-MOSCOW~}

\begin{document}

\noindent
 {\bf Corresponding author}\\
Prof. Dr. H.V. Klapdor-Kleingrothaus\\
Max-Planck-Institut f\"ur Kernphysik\\
Saupfercheckweg 1\\
D-69117 HEIDELBERG\\
GERMANY\\
Phone Office: +49-(0)6221-516-262\\
Fax: +49-(0)6221-516-540\\
email: $klapdor@gustav.mpi-hd.mpg.de$\\

\begin{center}
{\Large \bf 
SUPPORT OF EVIDENCE FOR NEUTRINOLESS DOUBLE BETA DECAY
} \\
\vspace{1.5cm}
{ H.V. Klapdor-Kleingrothaus
\footnote{Spokesman of HEIDELBERG-MOSCOW and GENIUS Collaborations,\\
E-mail: klapdor@gustav.mpi-hd.mpg.de,\\ 
 Home-page: $http://www.mpi-hd.mpg.de.non\_acc/$}${}^,$${}^2$, A. Dietz${}^2$,
I.V. Krivosheina${}^{2,3}$, Ch. D\"orr${}^2$ and C.~Tomei${}^{4}$}\\
{
\vspace{0.75cm}
{\sl ${}^2$Max-Planck-Institut f\"ur Kernphysik, P.O. 10 39 80}\\
{\sl D-69029 Heidelberg, Germany} \\
{\sl ${}^3$Radiophysical-Research Institute, Nishnii-Novgorod,
Russia}\\
{\sl ${}^4$Universita degli studi di L'Aquila, Italy}
}
\end{center}

\vspace{1.cm}

\begin{abstract}

	Indirect support for the evidence of neutrinoless double beta
	decay reported recently,  is obtained by analysis of other Ge double
	beta experiments, which yield independent information 
	on the background in
	the region of $Q_{\beta\beta}$. Some statistical features 
	as well as background
	si\-mulations with GEANT 4 of the HEIDELBERG-MOSCOW expe\-riment are
	discussed which disprove recent criticism.

\end{abstract}

\newpage

	Recently first experimental evidence has been reported 
	for neutrinoless
	do\-uble beta decay. Analysis of 55 kg y of data, taken by the
	HEIDELBERG-MOSCOW experiment in the GRAN SASSO over 
	the years 1990 - 2000, has led 
\cite{evid1,evid2,evid3,evid4} 
	to a half-life

\beq
T_{1/2} = (0.8-18.3)\times10^{25} \quad\mbox{years}\quad  (95\% \;C.L.),
\eeq

with best value of $T_{1/2}=1.5\times 10^{25}$ y,
for the decay of the double beta emitter $^{76}$Ge

\beq
^{76}\mbox{Ge} \longrightarrow ^{76}\mbox{Se} +2e^-.
\eeq

	Assuming the decay amplitude to be dominated by exchange of a massive
	Majorana neutrino (see, e.g. \cite{60years}), 
	this half-life results in a value
	of the effective neutrino mass

\beq
<m> = \left|  \sum U^2_{ei} m_i\right| = 0.05 - 0.84\; \mbox{eV} \quad(95\% \;C.L.),
\eeq

	with best value of 0.39 eV.
	Here a 50\% uncertainty in the nuclear matrix elements 
	has been taken into
	account (for details see 
\cite{evid3}).


\begin{figure}[th]
\begin{center}
\begin{sideways}
\begin{sideways}
\begin{sideways}
\includegraphics[width=7.0cm]{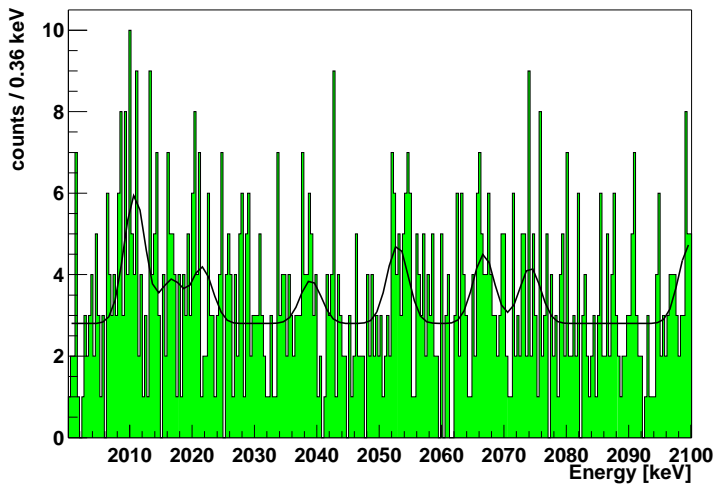}
\end{sideways}
\end{sideways}
\end{sideways}
\end{center}
\caption{\label{picMany}\rm \small 
	The spectrum taken with $^{76}$Ge detectors 
	Nrs. 1,2,3,4,5 over the period August 1990 - May 2000 (54.98 kg y), 
	in the energy range 2000 - 2100\,keV. 
	Simultaneous fit of the $^{214}$Bi-lines and the two high-energy
	lines yield a probability for a line at 2039 keV of 91\% C.L.}
\end{figure}


\begin{figure}[th]
\begin{center}
\includegraphics[width=6cm]{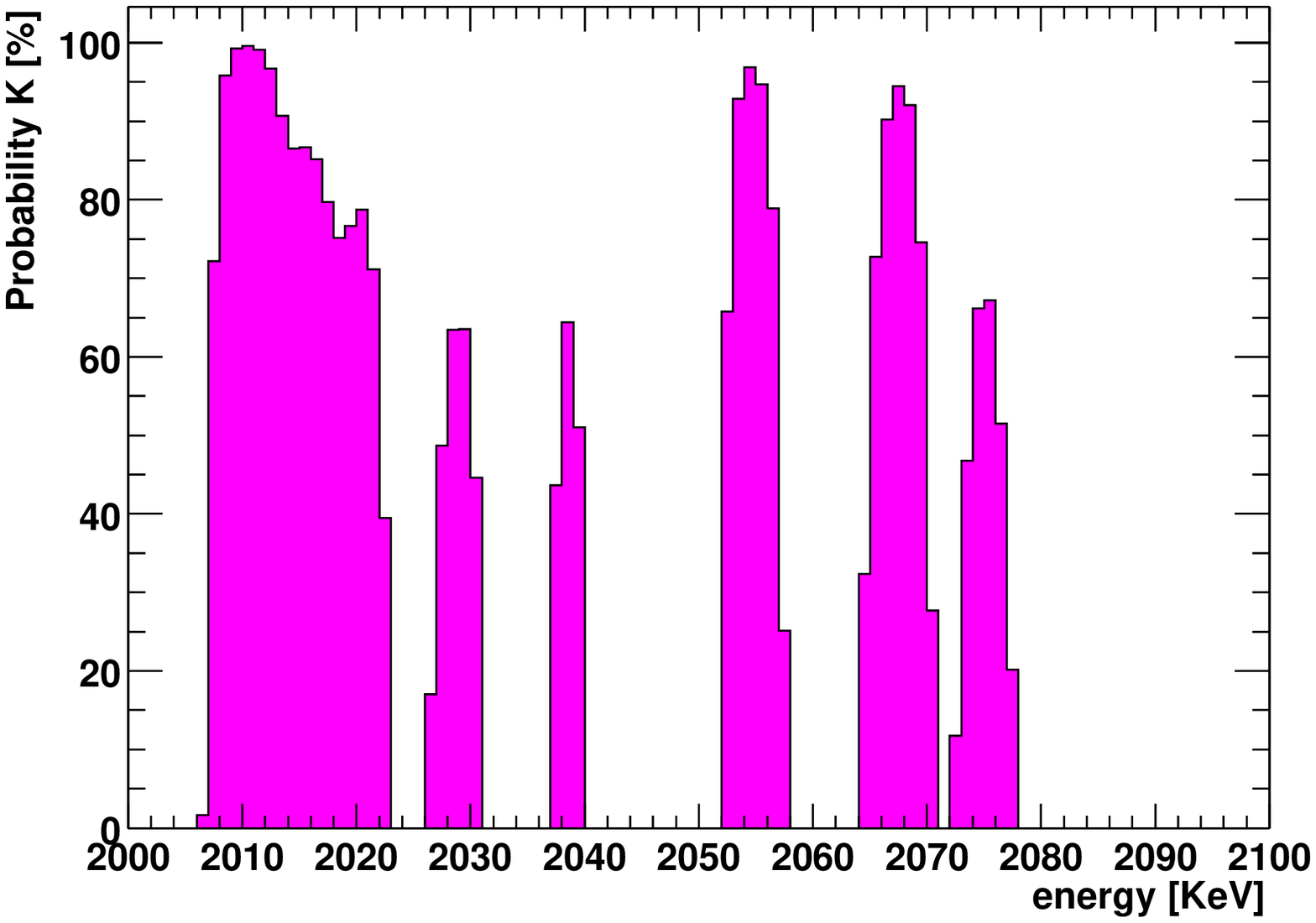}
\vspace*{0cm}
\includegraphics[width=6cm,height=4.5cm]{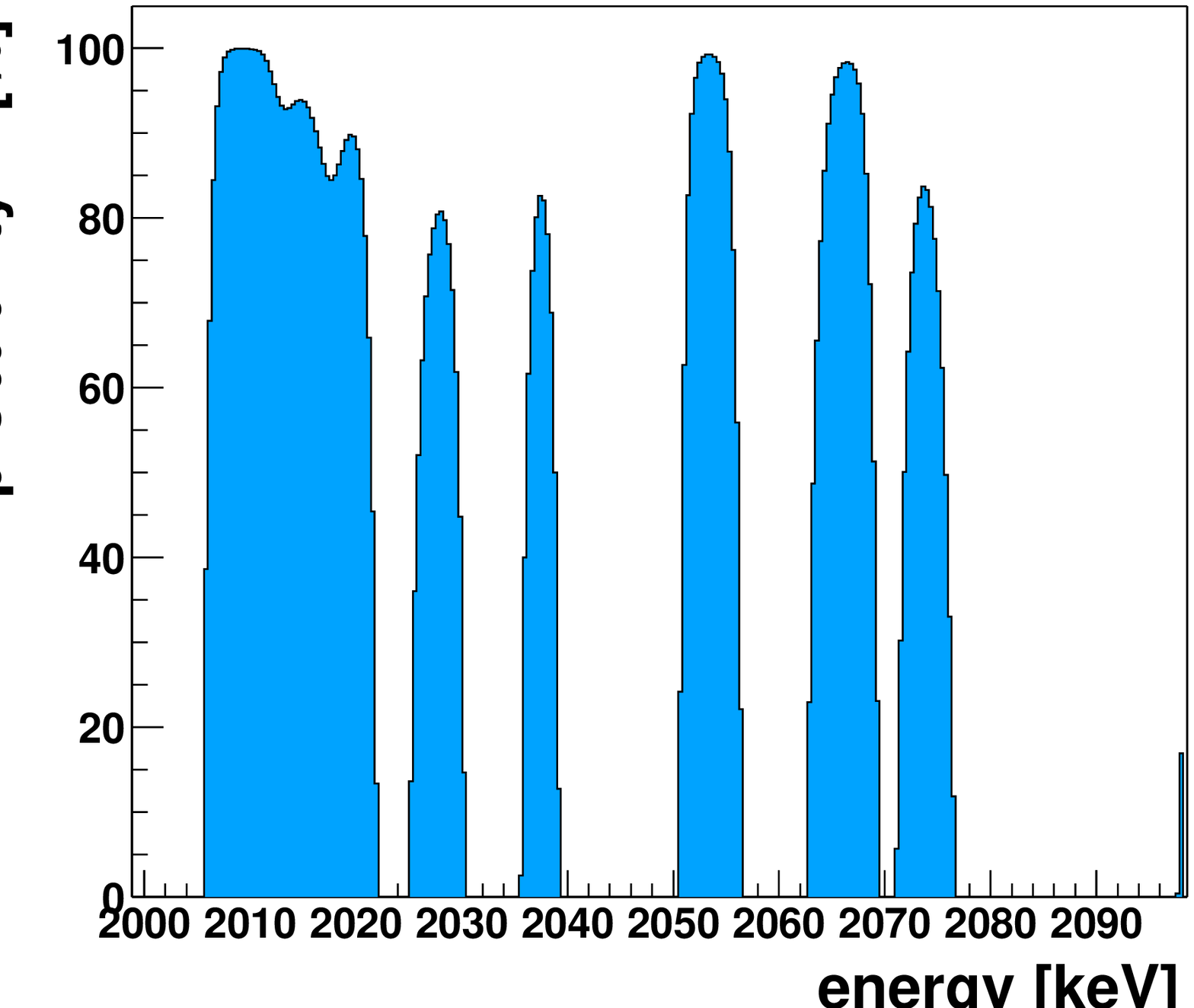}
\caption{\label{figHM}\rm \small 
	Result of the peak-searching procedure
	performed on the \HM spectrum (taken with detectors 1,2,3,5) 
	using the Ma\-ximum
	Likelihood approach (left) and the Bayesian method (right).
	On the y axis the probability 
	of having a line at the corresponding energy 
	in the spectrum is shown.}
\end{center}
\end{figure}


\begin{figure}[h!]
\begin{center}
\includegraphics[width=6.0cm]{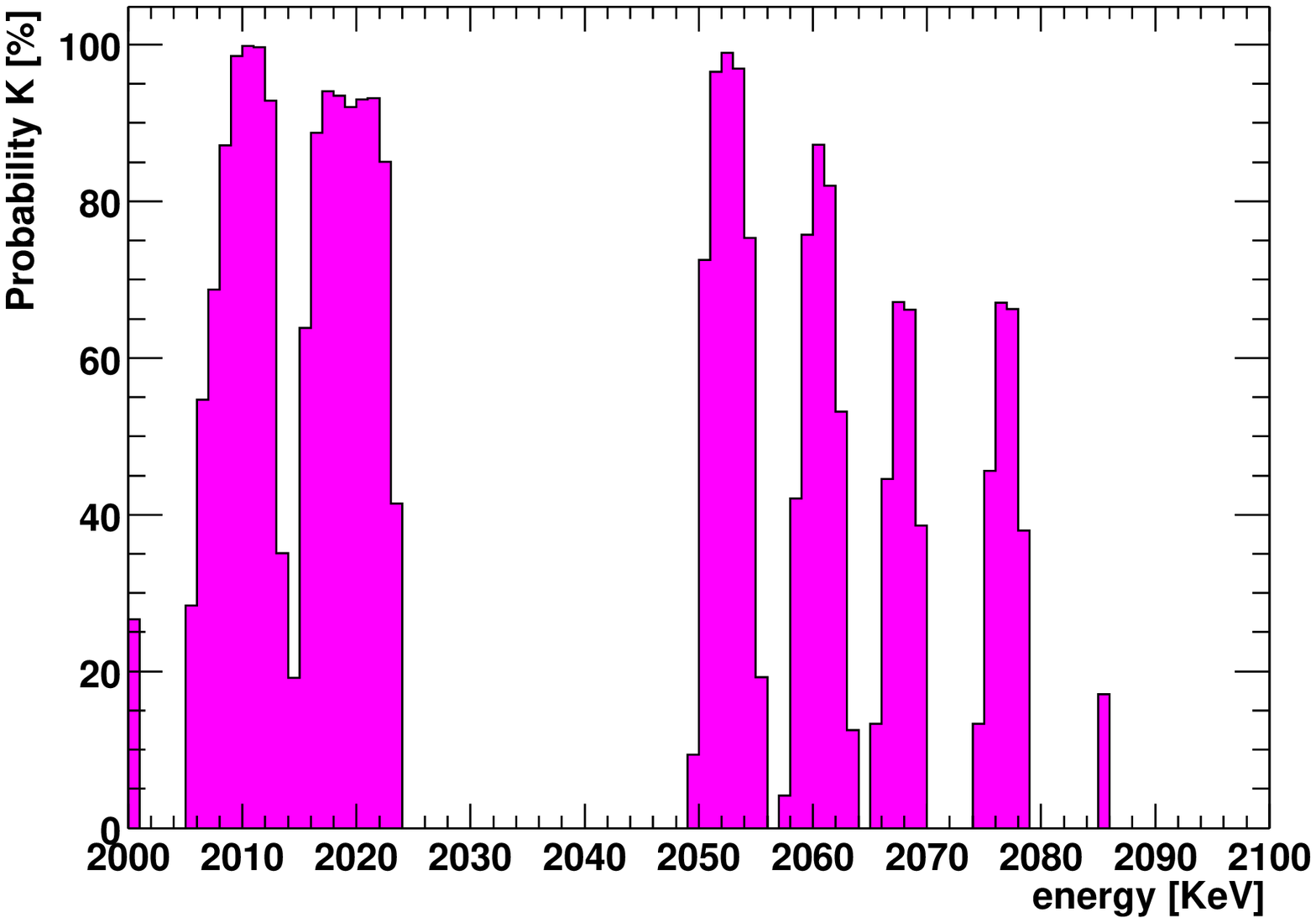}
\includegraphics[width=6.0cm]{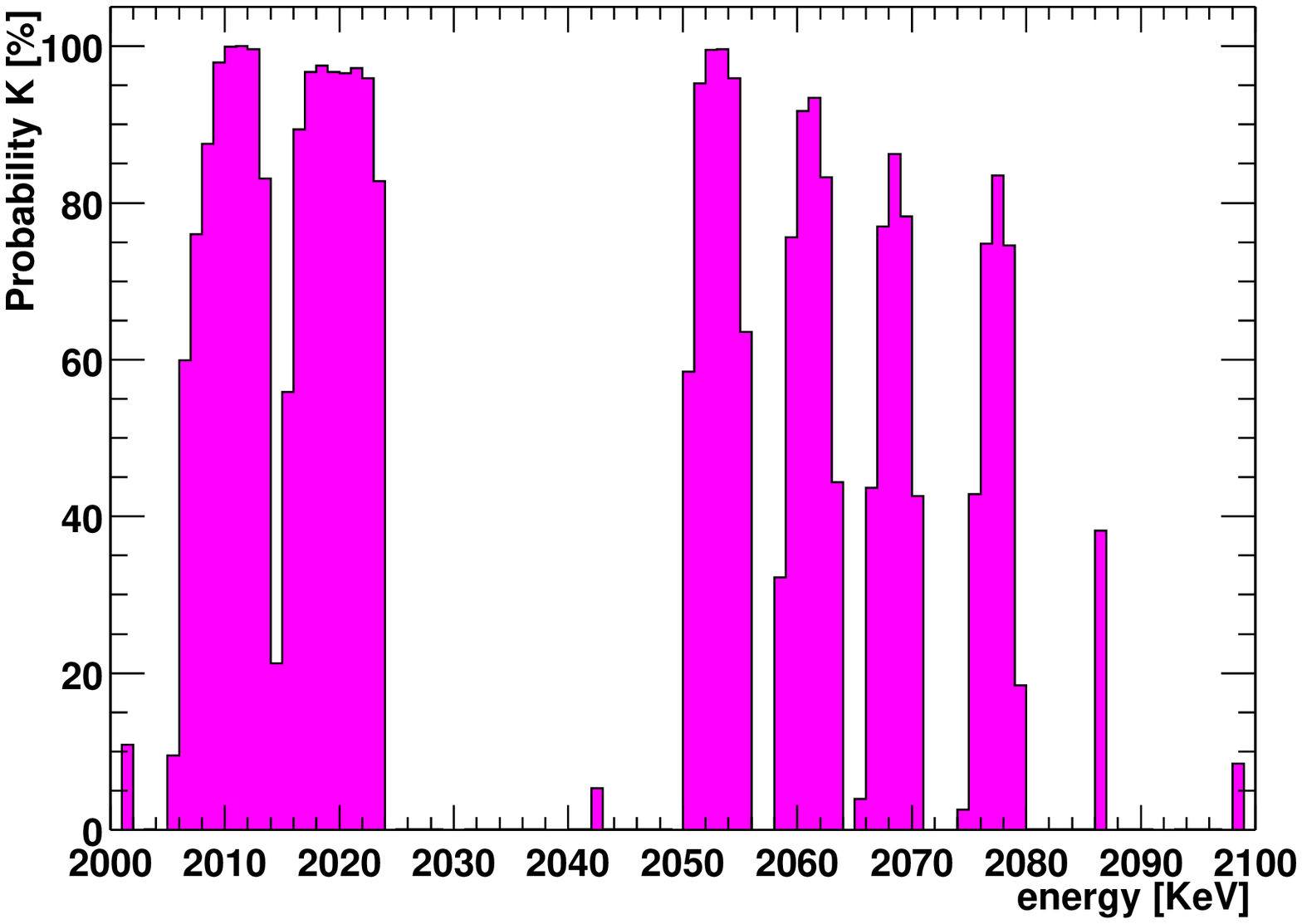}
\caption{\label{figCaldwell}\rm \small 
	Result of the peak-search procedure
	performed for the UCBS/LBL spectrum 
\cite{caldwell} 
	(left: Maximum Likelihood method, right: Bayes method). 
	On the y axis the probability 
	of having a line at the corresponding energy in the spectrum is shown.}
\end{center}
\end{figure}


	This is for the first time that the absolute scale 
	of the neutrino mass
	spectrum has been fixed, which cannot be achieved 
	by neutrino oscillation experiments. 
	This result restricts possible neutrino mass scenarios to
	degenerate or (still marginally allowed) inverse hierarchy
\cite{con1,con2,con3}. 
	In the degenerate case it leads to a common
	neutrino mass eigenvalue of

\beq
m_1 = 0.05 - 3.4\; \mbox{eV}  \quad(95\%\; C.L.).
\label{eq1}
\eeq

	This result is nicely consistent with later collected or analyzed
	experimental data, such as results from Large Scale Structure
	and CMB measurements 
\cite{cmb1,cmb2,cmb3}, 
	or ultra-high cosmic rays 
\cite{cray}.
	The former yield an upper limit of $\sum_i m_i$=1.0 eV 
	(corresponding in the degenerate case to a common 
	mass eigenvalue $m_0 < 0.33$ eV).
	The Z-burst scenario for ultra-high cosmic 
	rays requires 0.1 - 1.3\,eV  
\cite{cray}. 
	Tritium single beta decay cuts the upper range in 
eq.(\ref{eq1})  
	down to 2.2 or 2.8~eV 
\cite{tritium}.

	There is further theoretical support for a neutrino mass in the range
	fixed by the HEIDELBERG-MOSCOW experiment. 
	A model basing on an A4 symmetry
	of the neutrino mass matrix requires the
	neutrinos to be degenerate and the common mass eigenvalue to be
	$>$0.2~eV 
\cite{ma}. 
	Starting with the hypothesis that quark and lepton mixing are
	identical at or near the GUT scale, Mohapatra \etal  
\cite{mohap} 
	show that the large solar and atmospheric neutrino mixing 
	angles can be
	understood purely as result of renormalization group evolution, if
	neutrino masses are quasi-degenerate (with same CP parity).

	The common Majorana neutrino mass then must be larger than 0.1\,eV.
	The fact that WMAP and less stringent tritium decay cuts 
	away the upper part of the allowed range from double 
	beta decay (eq. (4)), may indicate 
\cite{KK-OULU02}
	that indeed 
	the neutrino mass eigenvalues have the same sigen 
	of CP phases - as required by 
\cite{mohap}.


\begin{figure}[t]
\begin{center}
\includegraphics[width=8cm]{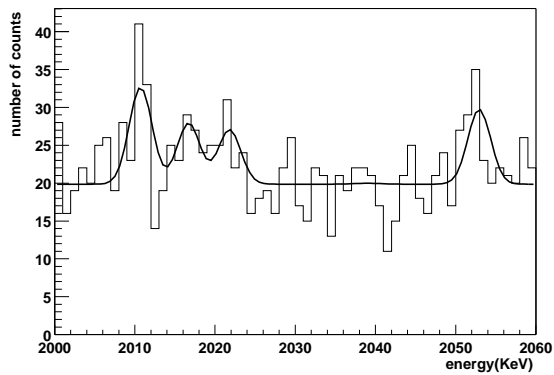}
\caption{\label{specCaldwell}\rm \small 
	Analysis of the spectrum measured by D. Caldwell et al. 
\cite{caldwell},
        with the Maximum Likelihood Method, in the energy 
        range 2000 - 2060\,keV 
        assuming lines at 2010.78, 2016.70, 2021.60, 2052.94,
	2039.0\,keV.        
        No indication for a signal at 2039\,keV is observed in this case.}
\end{center}
\end{figure}


	In this Letter we report additional support of the double beta 
	result of
\cite{evid1,evid2,evid3,evid4}, 
	by further discussion of the structure of the experimental
	background, by statistical considerations and by analysis 
	of other double beta experiments investigating 
	the decay of $^{76}$Ge.

	Important points in the analysis of the measured spectrum are the
	structure of the background around 
	$Q_{\beta\beta}$ (=2039.006(50) keV 
\cite{qvalue}), 
	and the energy range of analysis around $Q_{\beta\beta}$.

	{\it Background lines in the vicinity of $Q_{\beta\beta}$:} 
\hspace{0.2cm}
	Fig. 1 shows the spectrum measured in the range 2000 - 2100\,keV 
	in its original binning of 0.36\,keV. 
	By the peak search procedure developped 
\cite{evid2,evid3}
	on basis of the Bayes and Maximum Likelihood Methods, exploiting as
	important input parameters the experimental knowledge on the shape and
	width of lines in the spectrum, weak lines of $^{214}$Bi 
	have been identified
	at the energies of 2010.78, 2016.7, 2021.6 and 2052.94 keV 
\cite{evid1,evid2,evid3,evid4}.
	Fig. 2 shows the probability that there is a line 
	of correct width and of Gaussian shape at a given energy, 
	assuming all the rest of the spectrum as
	flat background (which is a highly conservative assumption).

	The intensities of these lines have been shown 
	to be consistent with other,
	strong Bi lines in the measured spectrum according 
	to the branching ratios
	given in the Table of Isotopes 
\cite{toi}, 
	and to Monte Carlo simulation of the
	experimental setup 
\cite{evid3}. 
	Note that the 2016 keV line, as an E0 transition,
	can be seen only by coincident summing of the two successive lines
	$E=1407.98$ keV
	and $E=609.316$ keV. Its observation proves that 
	the $^{238}$U impurity from which
	it is originating, is located in the Cu cap of the detectors.
	Recent measurements of the spectrum of a $^{214}$Bi {\it source } as
	function of distance source-detector confirm this interpretation 
\cite{oleg}.

	Premature estimates of the Bi intensities given in 
\cite{Dum-Comm01}
	thus are incorrect, because this long-known spectroscopic
	effect of true coincident summing 
\cite{gamma} 
	has not been taken into
	account, and also no simulation of the setup has been performed (for
	details see 
\cite{evid3,kk0205}). 

	These $^{214}$Bi lines occur also in other investigations 
	of double beta decay
	with Ge - and - even more important - also the additional 
	structures in
	Fig. 2, which cannot be attributed  at present, are seen in these
	other investigations.

	There are three other Ge experiments which have 
	looked for double beta
	decay of $^{76}$Ge. First there is the experiment by Caldwell et al.
\cite{caldwell}, 
	using natural Germanium detectors (7.8\% abundance of $^{76}$Ge, 
	compared to 86\% in the HEIDELBERG-MOSCOW experiment). 
	This was the most sensitive {\it natural} Ge
	experiment. With their background a factor of 9 higher than in the
	HEIDELBERG-MOSCOW experiment and their measuring time 
	of 22.6\,kg years,
	they had a statistics of the background by a factor of almost four
	\mbox{l a r g e r} than in the HEIDELBERG-MOSCOW experiment. 
	This gives useful
	information on the composition of the background.


\begin{figure}[th]
\begin{center}
\includegraphics[width=6.0cm]{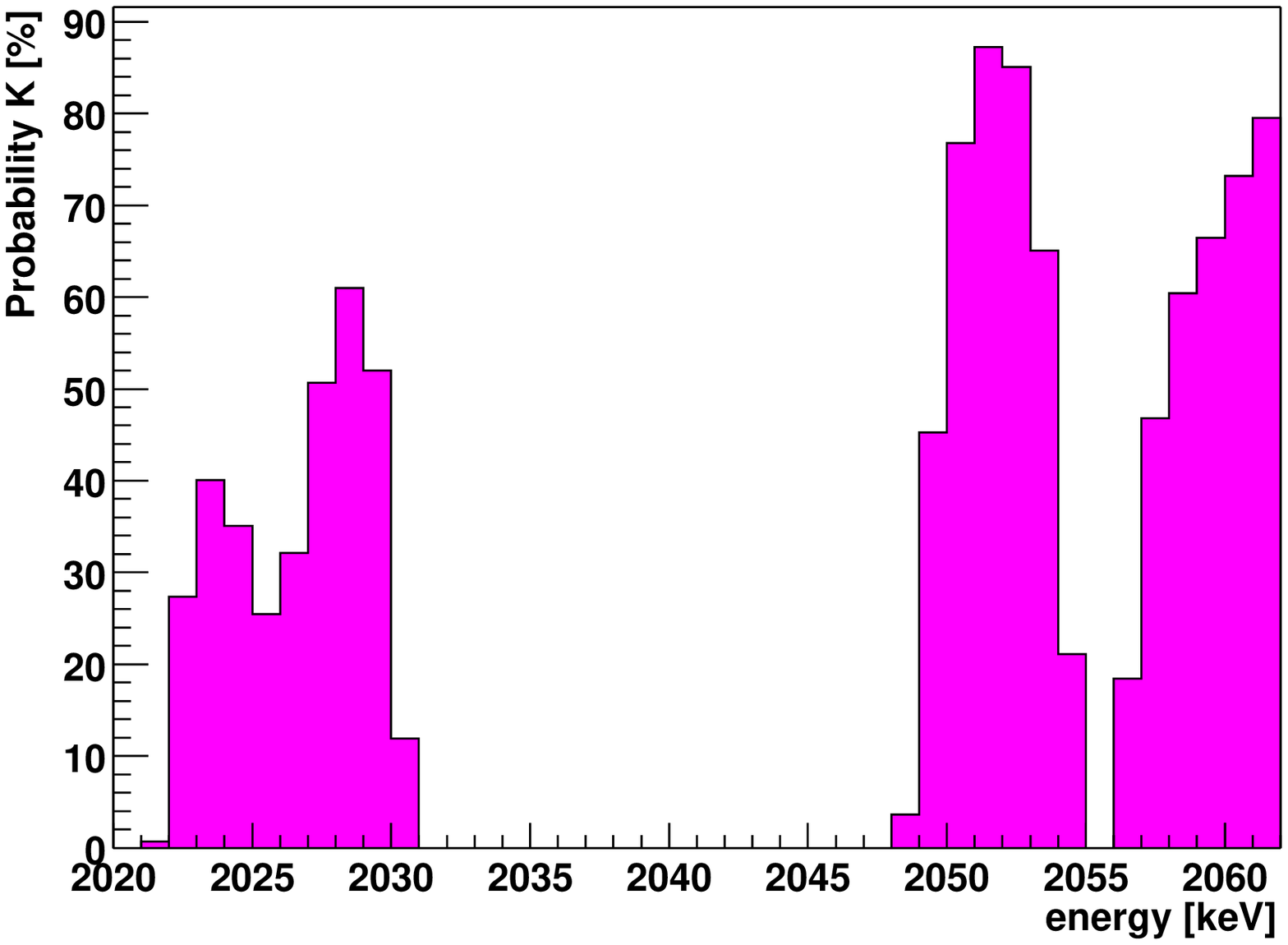}
\includegraphics[width=6.0cm]{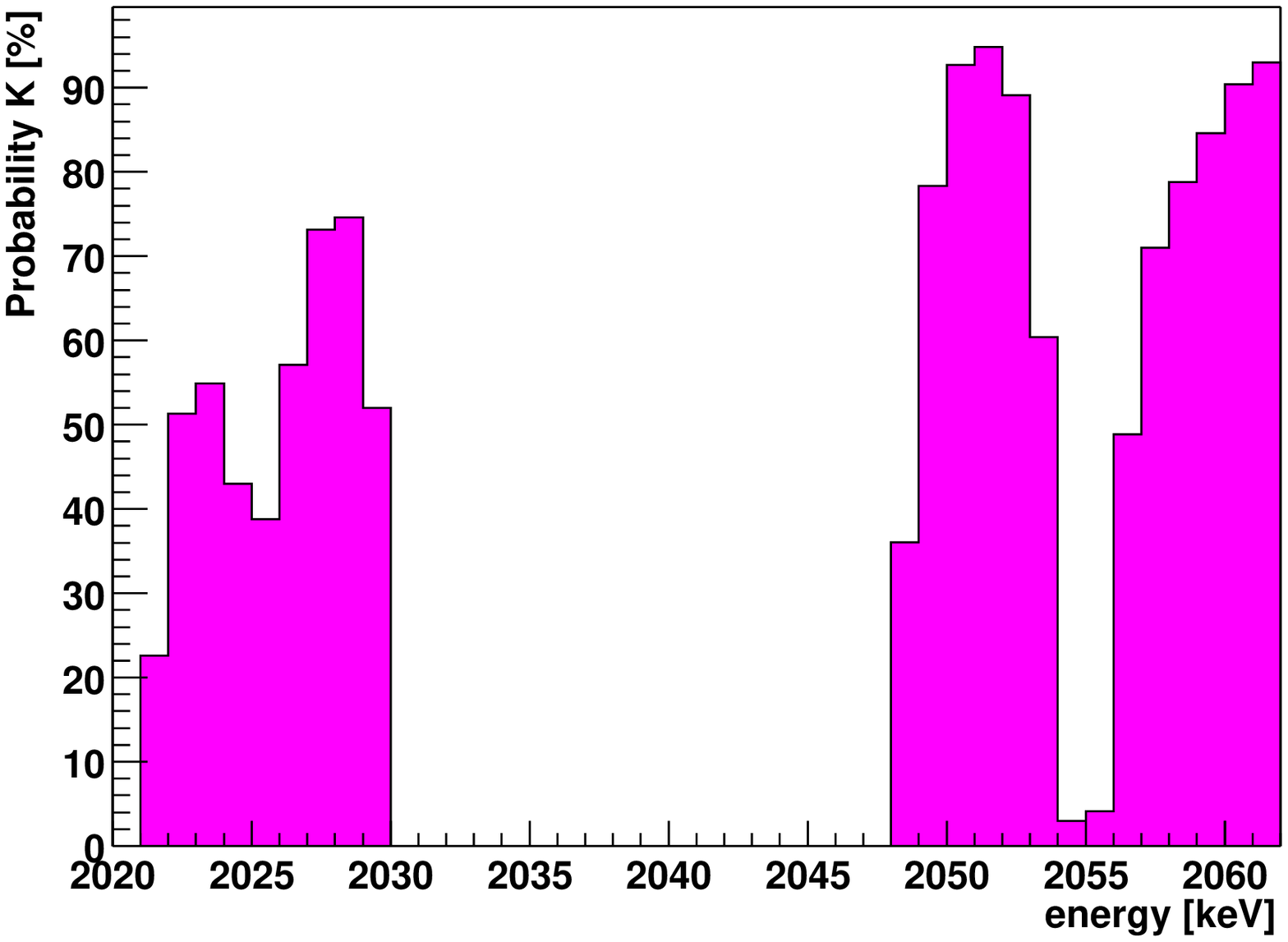}
\caption{\label{figITEP}\rm \small 
	Result of the peak-search procedure
	performed for the ITEP/YePI spectrum 
\cite{vasenko} 
	(left: Maximum Likelihood method, right: Bayes method). 
	On the y axis the probability 
	of having a line at the corresponding energy 
	in the specrtum is shown.}
\end{center}
\end{figure}



\begin{figure}[h!]
\begin{center}
\includegraphics[width=6.0cm]{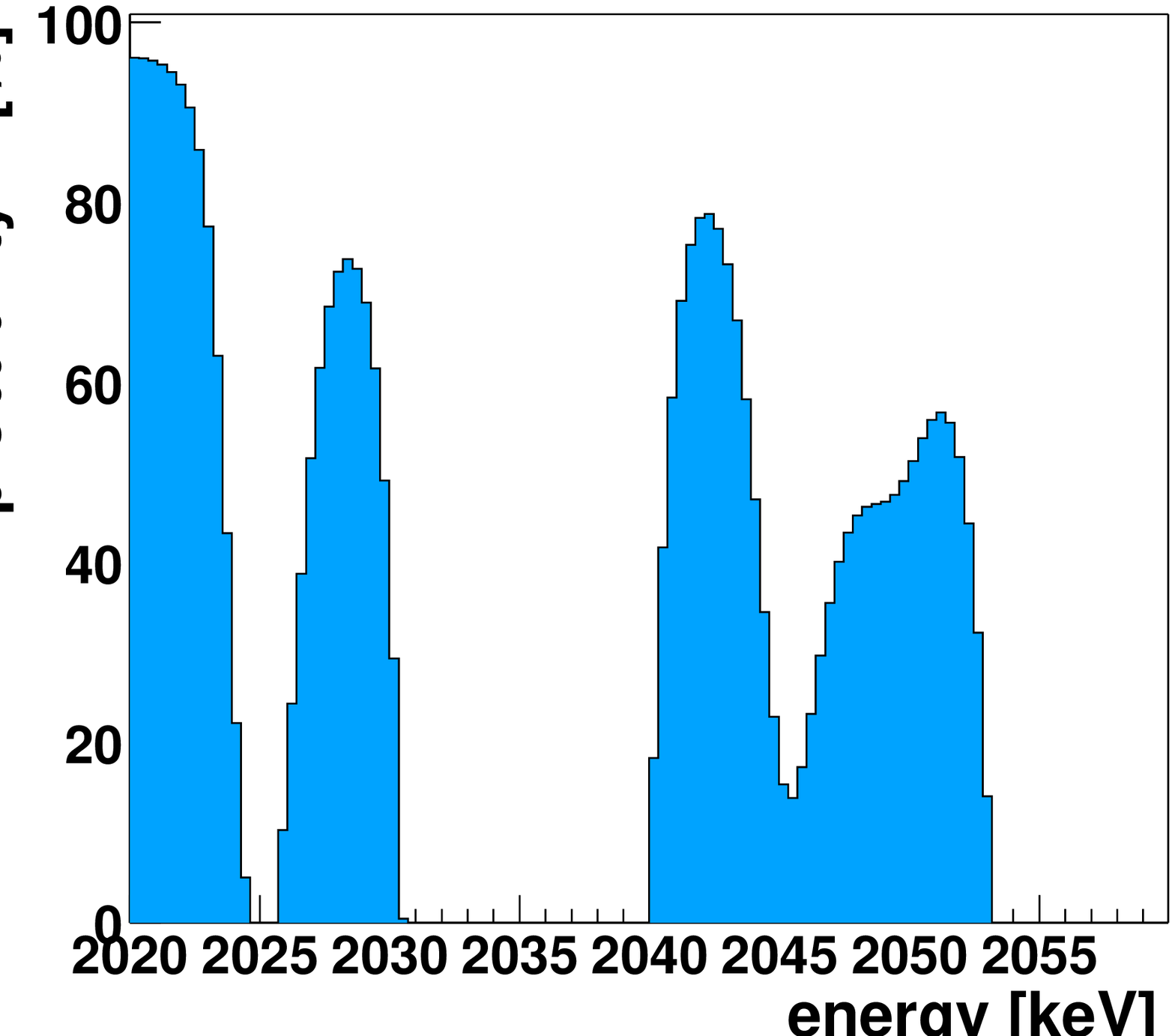}
\includegraphics[width=6.0cm]{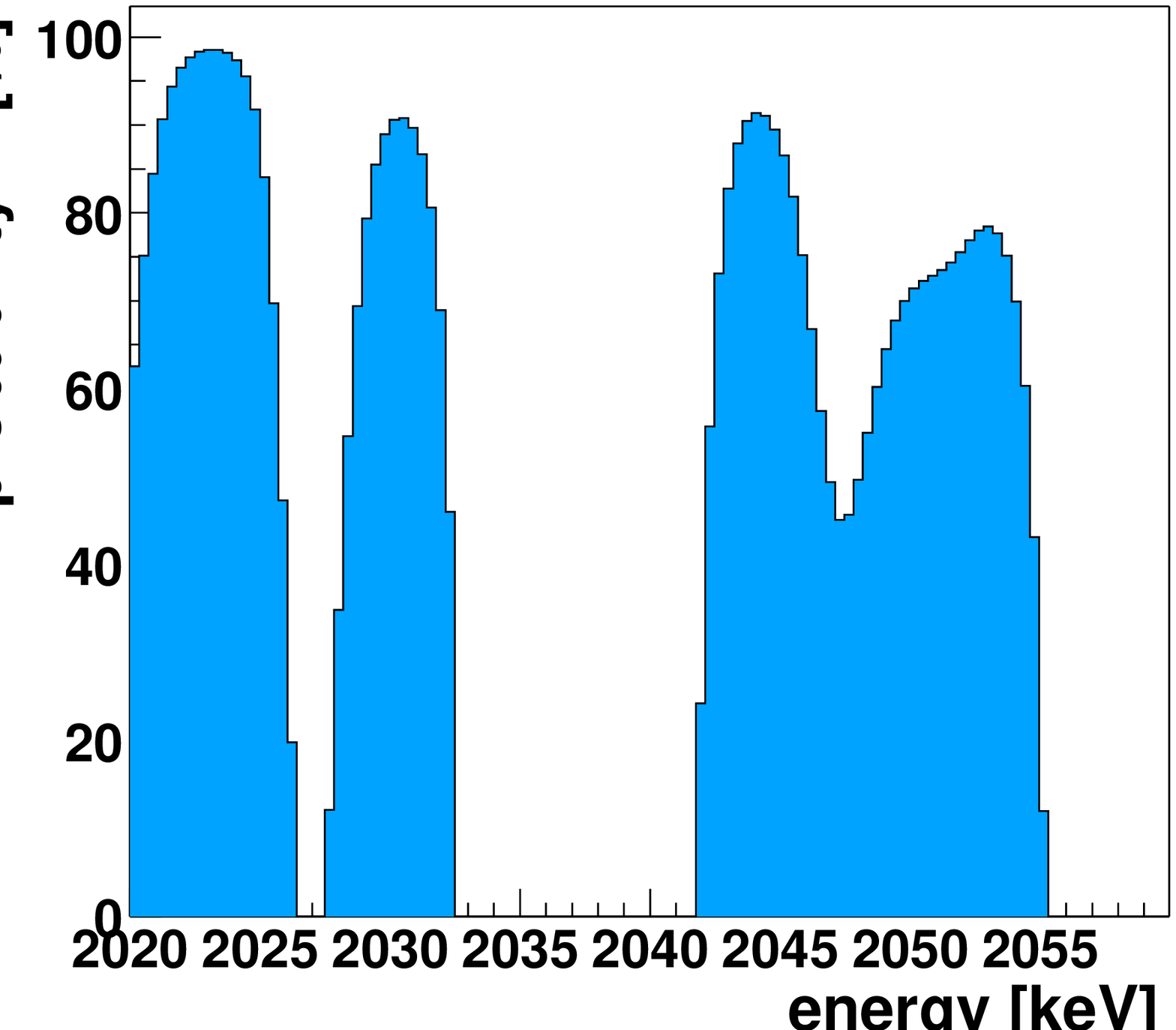}
\caption{\label{figIGEX}\rm \small 
	Result of the peak-search procedure
	performed for the IGEX spectrum 
\cite{igex,igex2} 
	using the ML approach
	(left) and the Bayesian statistics (right). 
	On the y axis the probability 
	of having a line at the corresponding energy 
	in the spectrum is shown.}
\end{center}
\end{figure}


	Applying the same method of peak search as used in Fig. 
\ref{figHM}, 
	yields indications for peaks essentially 
	at the same energies as in Fig. 
\ref{figHM} 
	(see Fig. 
\ref{figCaldwell}).  
	This shows that these peaks are not fluctuations. 
	In particular it sees the 2010.78, 2016.7, 2021.6 
	and 2052.94\,keV $^{214}$Bi lines, but
	a l s o  the unattributed lines at higher energies.  
	It finds, however,
	n o  line at 2039\,keV.  
	This is consistent with the
	expectation from the rate found in the HEIDELBERG-MOSCOW 
	experiment. About 16 observed events in the latter 
	correspond to 0.6  expected events in the Caldwell 
	experiment, because of the use of non-enriched
	material and the shorter measuring time. 
	Fit of the Caldwell spectrum
	allowing for the $^{214}$Bi lines and a 2039\,keV 
	line yields 0.4\,events for
	the latter (see Fig. 
\ref{specCaldwell}).

	The first experiment using enriched (but not high-purity) 
	Germanium 76
	detectors was that of Kirpichnikov and coworkers 
\cite{vasenko}. 
	These authors show only the energy range between 2020 and 2064\,keV of
	their measured spectrum.
	The peak search procedure finds also here indications of lines
	around 2028\,keV and 2052\,keV (see Fig. 
\ref{figITEP}), 
	but \mbox{n o t} any indication of a line at
	2039\,keV. This is consistent with the expectation, 
	because for their low
	statistics of 2.95\,kg\,y they would expect here (according to
	HEIDELBERG-MOSCOW) 0.9\,counts.

	Another experiment (IGEX) used between 6 
	and 8.8\,kg of enriched $^{76}$Ge,
	but collected since beginning of the experiment in the early nineties
	till shutdown in 1999 only 8.8\,kg\,years of statistics 
\cite{igex,igex2}.
	The authors of 
\cite{igex,igex2} 
	unfortunately show only the range
	2020 to 2060\,keV of their measured spectrum in some detail. Fig. 
\ref{figIGEX}  
	shows the result of our peak scanning 
	of this range. Clear indications are seen for
	the Bi lines at 2021 and 2052\,keV, but also 
	(as this is the case for the spectrum of 
\cite{vasenko}, 
	see Fig.5 ) 
	of the unidentified 
	structure around 2030\,keV. Because of the conservative 
	assumption on the background treatment in the scanning 
	procedure (see above) there is no chance to see
	a signal at 2039\,keV because of the 'hole' in the background 
	of that spectrum (see Fig. 1 in 
\cite{igex}). 
	With some good will one might see, however,
	an indication of 3 events here, consistent with the expectation of the
	HEIDELBERG-MOSCOW experiment of 2.6\,counts.


\begin{figure}[h]
\begin{center}
\includegraphics[width=6.7cm]{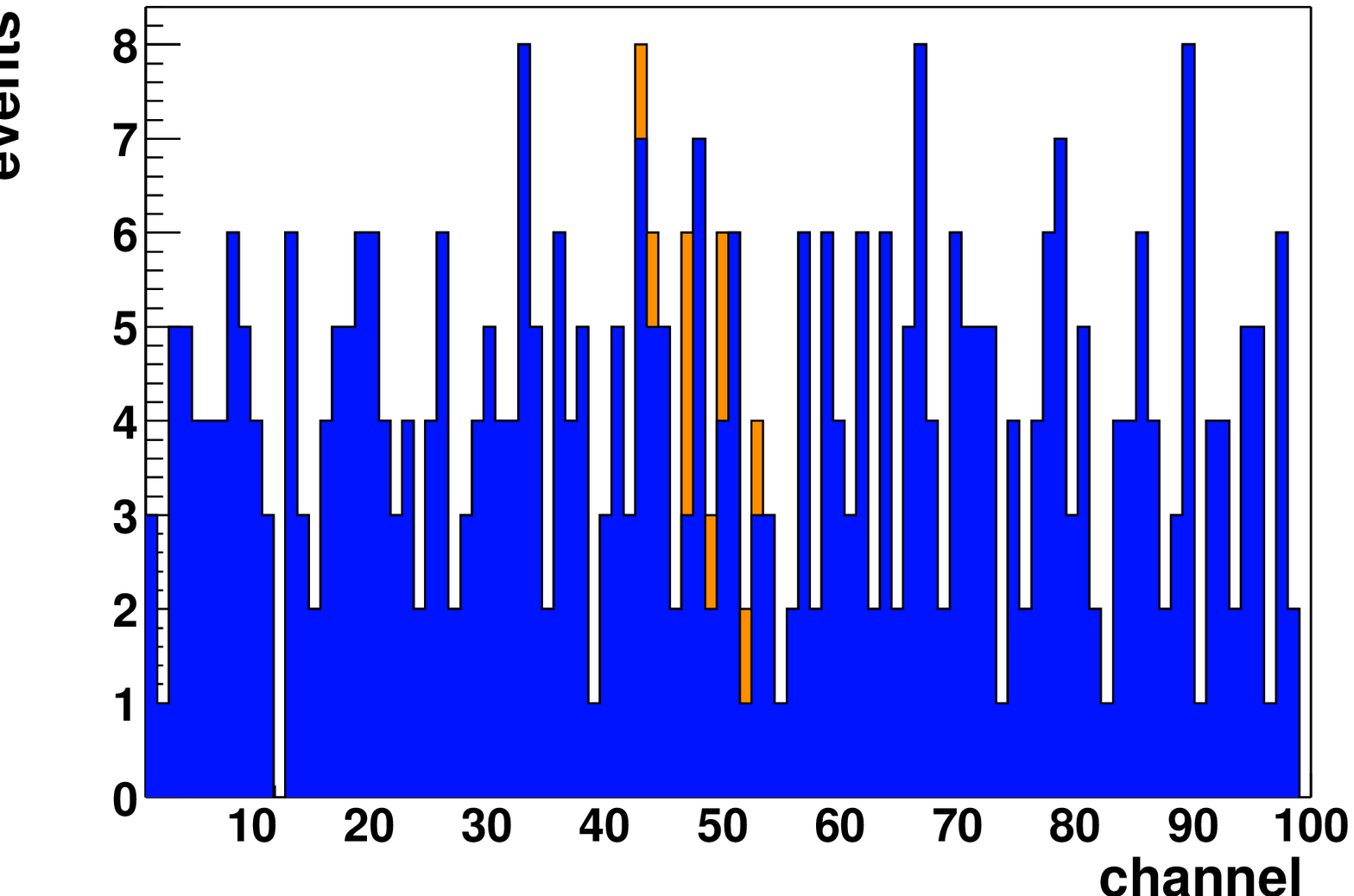}
\includegraphics[width=6.7cm]{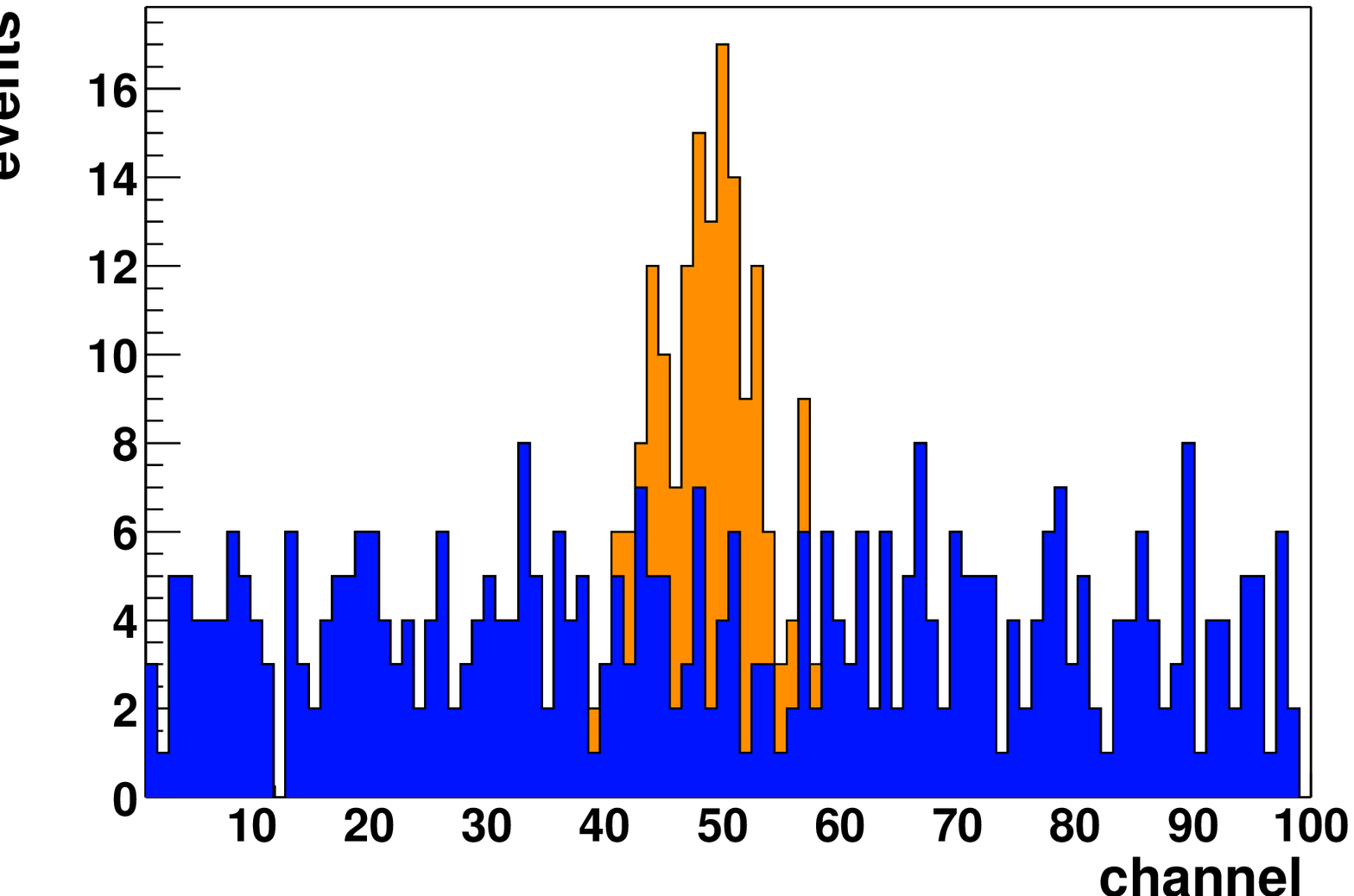}
\end{center}
\caption{\label{picSpecKu}\rm \small 
	Example of a random-generated spectrum with 
	a Poisson distributed
	background with 4.0 events per channel and a Gaussian 
	line centered in channel
	50 (line-width corresponds to a standard-deviation of $\sigma=4.0$
	channels).
	The left picture shows a spectrum with a line-intensity of 10 events,
	the right spectrum a spectrum with a line-intensity of 100 events.
	The background is shown dark, the events of the line bright.}
\end{figure}



\begin{figure}[th]
\begin{center}
\includegraphics[width=6.7cm]{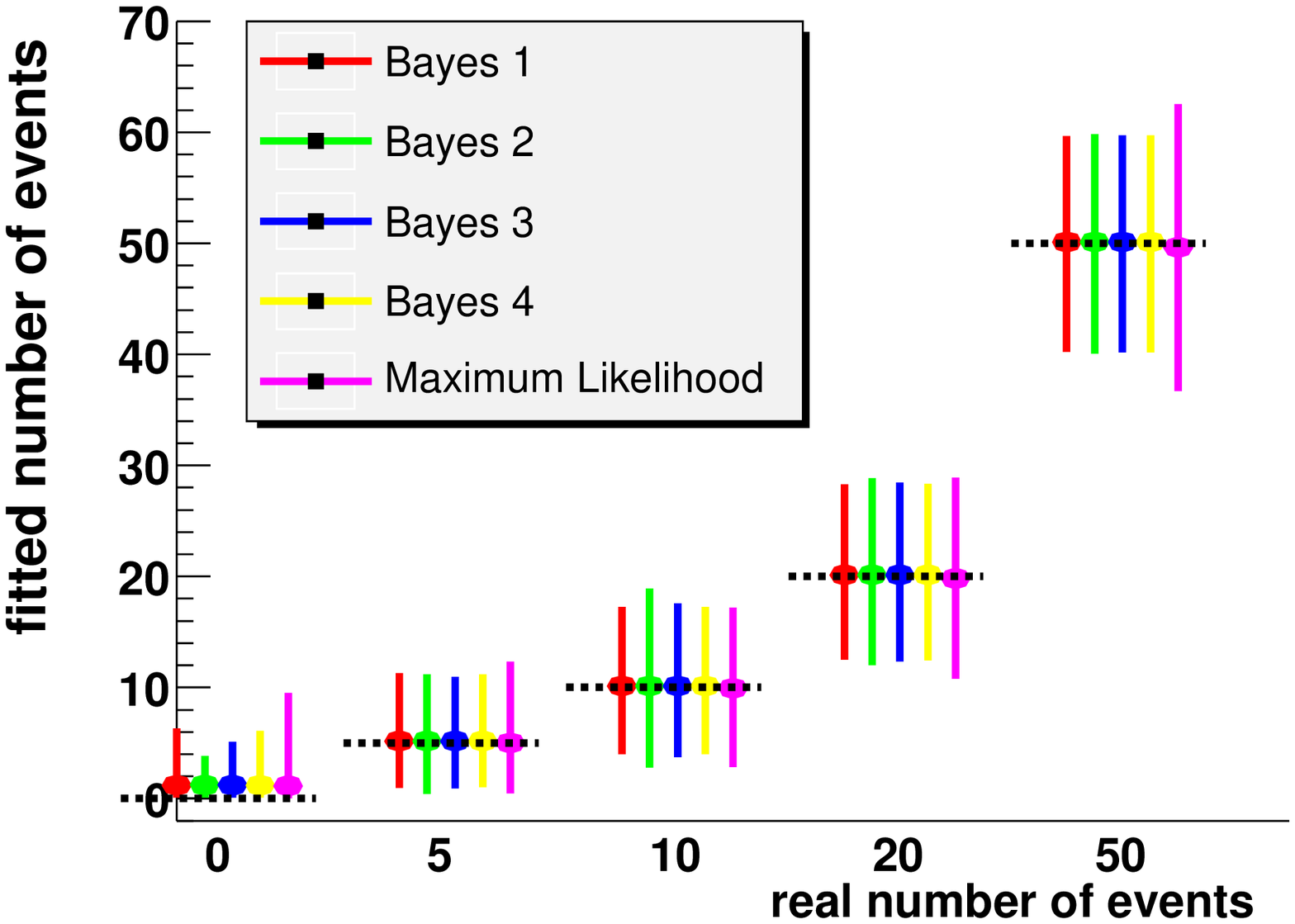}
\includegraphics[width=6.7cm]{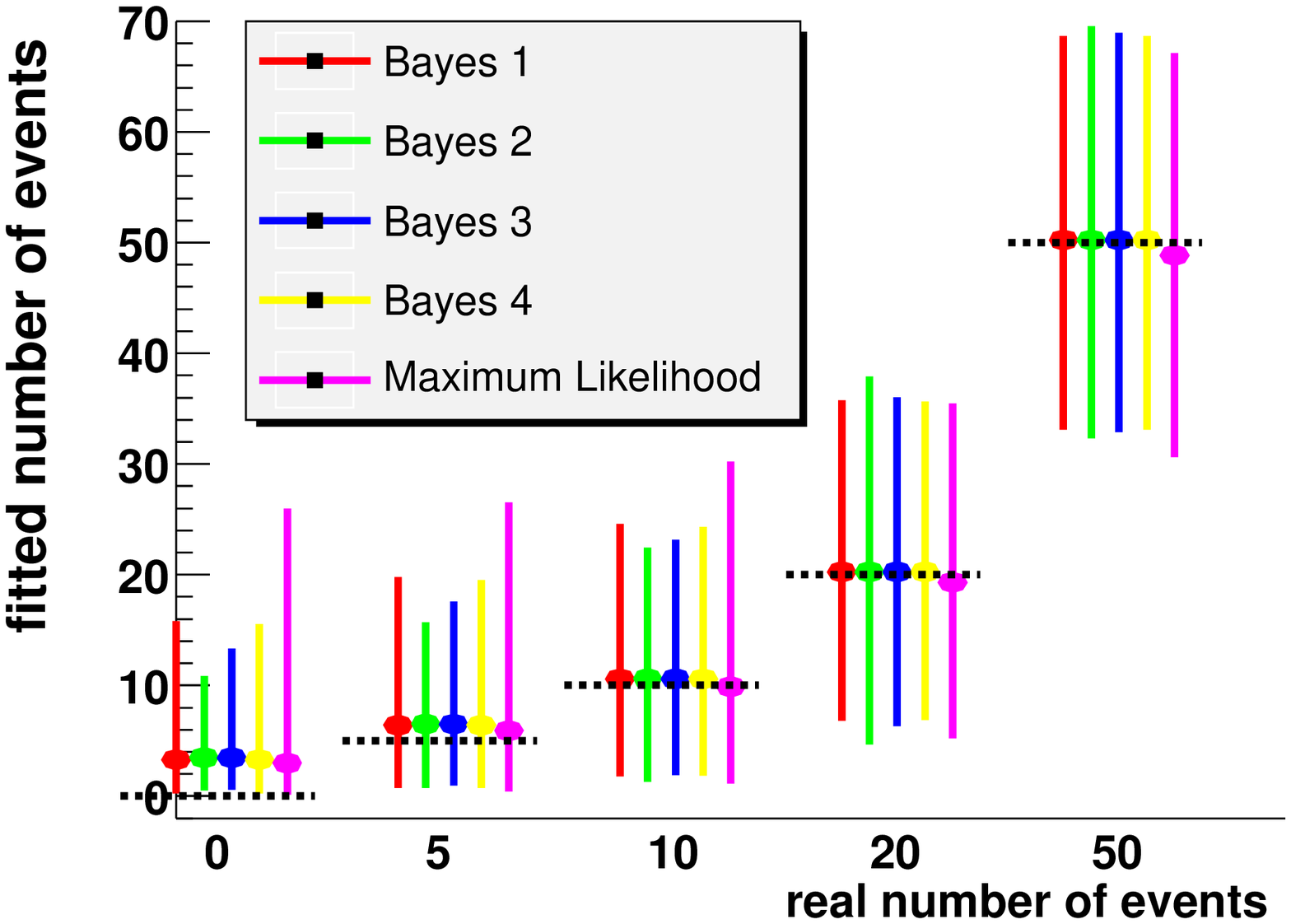}
\end{center}
\caption{\label{picPrior} \rm \small 
	Results of analysis of random-number generated spectra, using Bayes
	and Maximum Likelihood method (the first one 
	with different prior distributions).
	For every number of events in the simulated line, 
	shown on the x-axis, 1000 random generated 
	spectra were evaluated with the five given methods.
	The analysis on the left side was performed with an Poisson
	distributed background of 0.5 events per channel, the background for
	the spectra on the right side was 4.0 events per channel.
	Every vertical line shows the mean value of the calculated best values
	(thick points) with the 1$\sigma$ error area.
		The mean values are in good agreement with 
	the expected values (black
dashed lines).}
\end{figure}


	{\it The power of the peak search procedure:} \hspace{0.2cm}
	At this point it may be useful to demonstrate the potential 
	of the used peak search procedure. 
	Fig. 
\ref{picSpecKu} 
	shows a spectrum with Poisson-generated background of
	4 events per channel and a Gaussian line with width (standard
	deviation) of 4 channels centered at channel 50, with intensity of 10
	(left) and 100 (right) events, respectively.
	Fig. 
\ref{picPrior}, 
	right shows the result of the analysis of spectra of
	different line intensity with the Bayes method
	(here Bayes 1-4 correspond to different choice 
	of the prior distribution:
	(1) $\mu(\eta)=1$ (flat), (2) $\mu(\eta) = 1/\eta$, 
	(3) $\mu(\eta) = 1/\sqrt{\eta}$,
	(4) Jeffrey's prior) and the Maximum Likelihood Method.
	For each prior 1000 spectra have been generated with 
	equal background and equal line intensity using random 
	number generators available at CERN 
\cite{random}.
	The average values of the best values agree (see Fig. 
\ref{picPrior}) 
	very well with the known intensities also for very low 
	count rates (as in Fig.
\ref{picSpecKu}, 
	left).

\begin{figure}[h!]

\vspace{-0.5cm}
\begin{center}
\includegraphics[width=9cm]{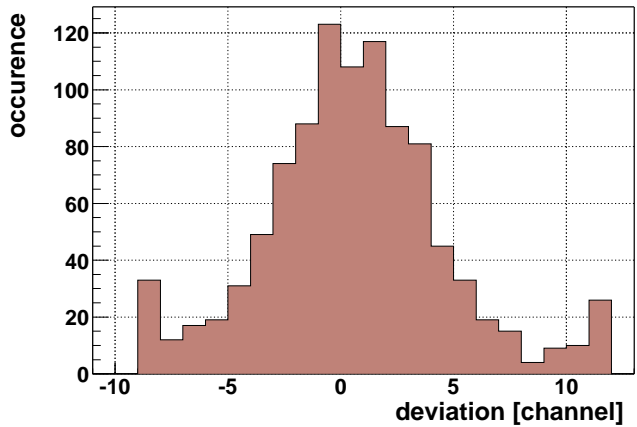}
\end{center}
\caption{\label{evalScan1} \rm \small 
	Simulated spectra with a Poisson-distributed background 
	and a Gausian line with 15\,events centered at channel 50 
	with a width (standard deviation) of 4.0 channels, 
	created with different random numbers. 
	Shown is the distribution of the deviation of the real 
	position of the lines, obtained by the peak scan procedure, 
	for 1000 spectra. The rms value of this distribution is 4.00 
        channels, corresponding to 1.44 keV in the spectra of the 
        HEIDELBERG-MOSCOW experiment. }
	
\end{figure}



\begin{figure}[h!]
\begin{center}
\includegraphics[width=10cm]{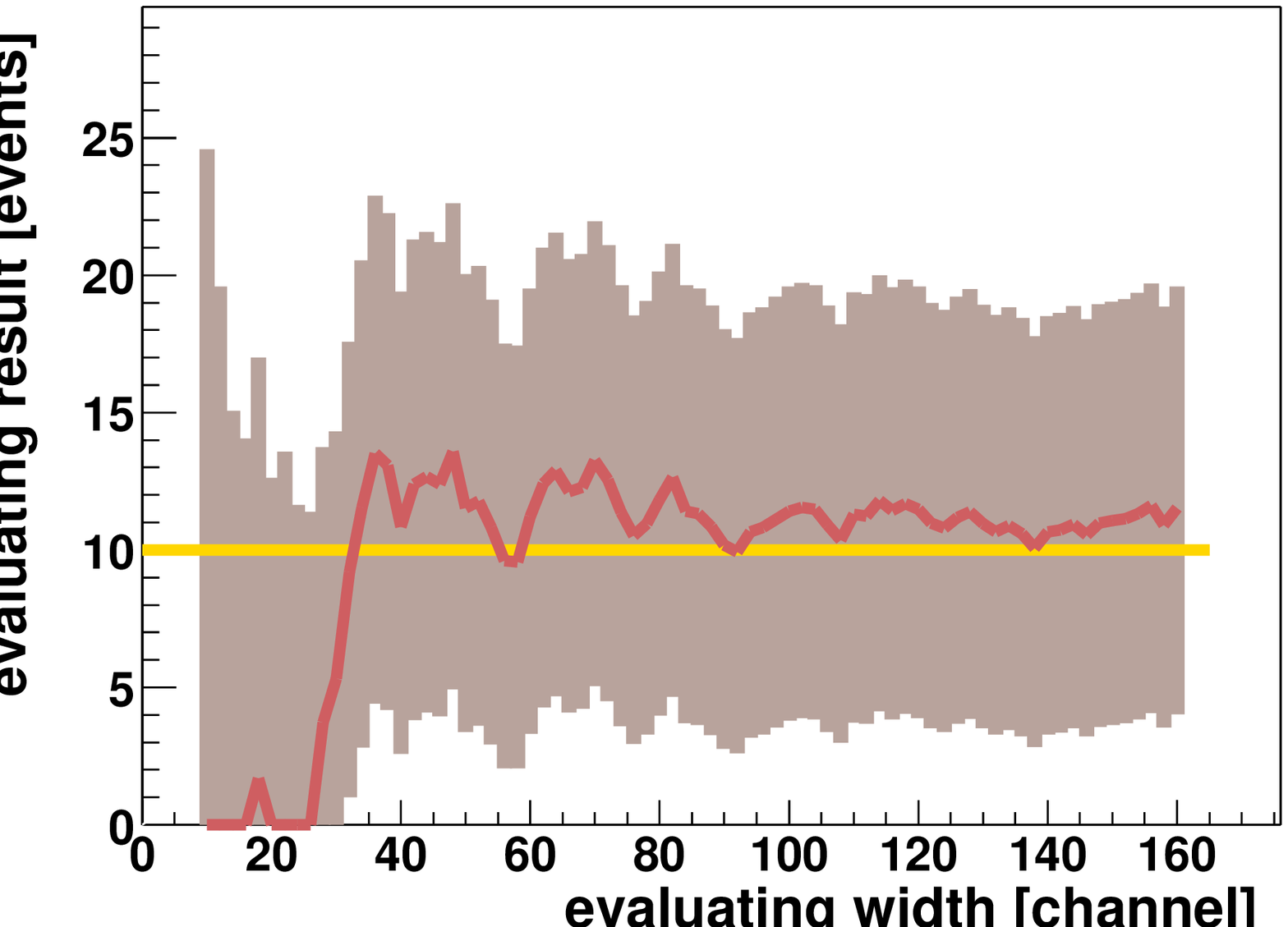}
\end{center}
\caption{\label{picSingleDev} \rm \small 
	Result of an analysis as function of the evaluation width.
	The used spectrum consists of a Poisson distributed background with 4
	events per channel, and a line of 10 events (see fig. 
\ref{picSpecKu},
	left part). 
	The dark area corresponds to a 68.3\% confidence area with the dark
	line being the best value.
	Below an evaluation width of 35 channels the result becomes unreliable,
	above 35 channels the result is stable.}
\end{figure}


	A detailed simulation has been done to show the behavior 
	of the position of the peak maximum of 
	a line of 15\,events, with width (standard deviation) of 4 channels, 
	on a Poisson-distributed background with 0.5\,events/channel.
	For that reason, a peak scan procedure was performed on 1000 randomly 
	created spectra.
	For each spectrum, the deviation of the real position of the line 
	to the position of the maximum probability was determined.
	The distribution of these deviations is shown in figure 
\ref{evalScan1}, 
	which gives a RMS value of 4.00 channels, corresponding 
	to $\sim$ 1.44 keV in the spectra from 
	the HEIDELBERG MOSCOW experiment.
	The figure
\ref{evalScan1} 
	describes the possible degree of deviation of the energy of the
	peak maximum from the transition energy,  on the level of statistics
	collected in experiments like the HEIDELBERG-MOSCOW experiment. 
	This should be considered when comparing
		Figs. 
\ref{figHM},\ref{figCaldwell},\ref{figITEP},\ref{figIGEX}.

\begin{figure}[t]
\begin{center}
\includegraphics[width=8.cm]{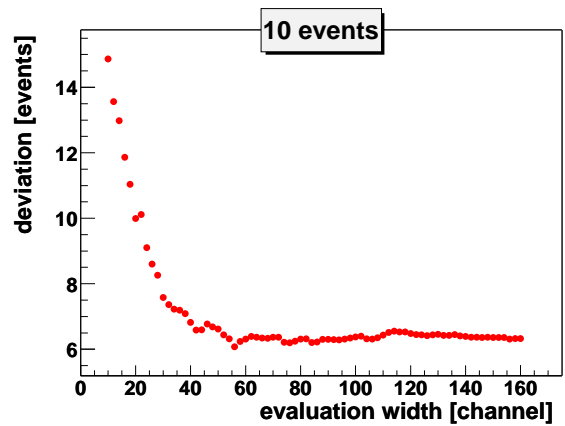}
\caption{\label{WiDev}\rm \small 
	Simulation of spectra with a line intensity of 10\,events for 
	a Poisson-distributed background with 4.0\,events/channel. 
	The horizontal axis shows the energy range of analysis. 
	The vertical axis shows the mean deviation of the intensities 
	obtained by the peak scanning procedure from the known 
	real intensities. Each dot represents the mean deviation 
	obtained from 100\,random spectra. The deviation is constant, 
	when more than 35\,channels are used for the evaluation.}
\end{center}
\end{figure}

	{\it Influence of the choice of the energy range of analysis:}
	\hspace{0.2cm}
	The influence of the choice of the energy range 
	of the analysis around $Q_{\beta\beta}$
	has been thoroughly discussed in 
\cite{evid2,evid3}. 
	Since erroneous ideas about this
	point are still around, a few further comments may be given here. In
	Fig. 
\ref{picSingleDev} 
	we show the analysis of a simulated spectrum
	consisting of a Gaussian line of width (standard deviation) 
	of 4 channels and intensity of 10 counts 
	on a Poisson-distributed background of 4 events
	per channel (see fig. 
\ref{picSpecKu} 
	left), as function of the width of the range of analysis. 
	It is seen that a reliable result is
	obtained for a range of analysis of not smaller than 35 channels
	(i.e. $\pm$18 channels) - one channel corresponding to 0.36 keV in the
	HEIDELBERG-MOSCOW experiment. 
	Fig. 
\ref{WiDev} 
	shows this in a more general way. 
	Every dot 
	represents the mean deviation of the calculated line intensity 
	to the real line intensity, obtained from 100 random spectra. 
	The exact value for the deviation $\Delta$ is calculated by

$\Delta=\frac{1}{N} \sum\limits_{i=1}^{N} \left| n-b_i \right|$
$\quad N=100 $
$\quad n=0,5,10,20,50,$

	with $b_i$ being the result for the line intensity from the evaluation 
	and $n$ 
	the real line intensity, chosen as 10\,events in Fig. 
\ref{WiDev}.
	The spectra used here have again a background of 4.0 events 
	per channel.  
	Also in this general case it is seen that, 
	if the evaluation width is more than 35 channels, the deviation 
	is constant, so that an evaluation width of at least 35 channels 
	is required for reliable results.
	This is an important result, since it is of course important 
	to keep the range of analysis as  \mbox{s m a l l} as possible, 
	to avoid to include lines in the vicinity of the weak signal 
	into the background.
	This unavoidably occurs when e.g. proceeding as suggested in 
\cite{Dum-Comm01}.
	The arguments given in those papers are therefore incorrect. Also
	Kirpichnikov, who states 
\cite{kirch} 
	that his analysis finds 
	a 2039 keV signal in
	the HEIDELBERG-MOSCOW spectrum on a 4 sigma confidence level 
	(as we also see it, when using the Feldman-Cousins method 
\cite{dietzdiss}), 
	makes this mistake when analyzing the pulse-shape spectrum.


\begin{figure}[th]
\begin{center}
\includegraphics[width=14cm]{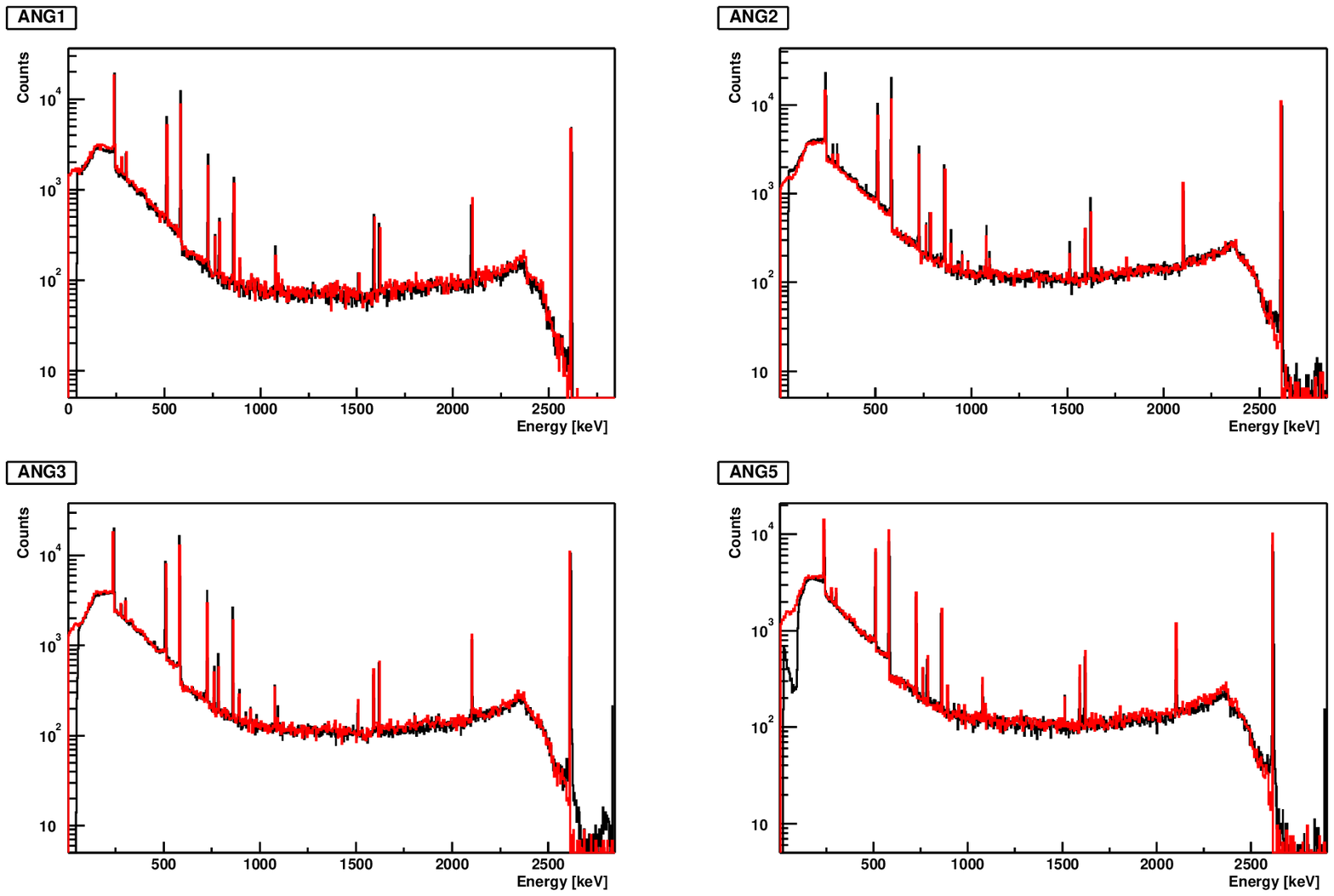}
\caption{\label{picUnderTotal}\rm \small 
	Comparison of the
	measured data (black line, November 1995 to April 2002) and simulated
	spectrum (red line) for the detectors Nrs. 1,2,3 and 5 for a
	$^{232}$Th source spectrum.
	The agreement of simulation and measurement is excellent.}
\end{center}
\end{figure}



\begin{figure}[tb]\vspace*{-0.5cm}
\begin{center}
\includegraphics[width=10.0cm]{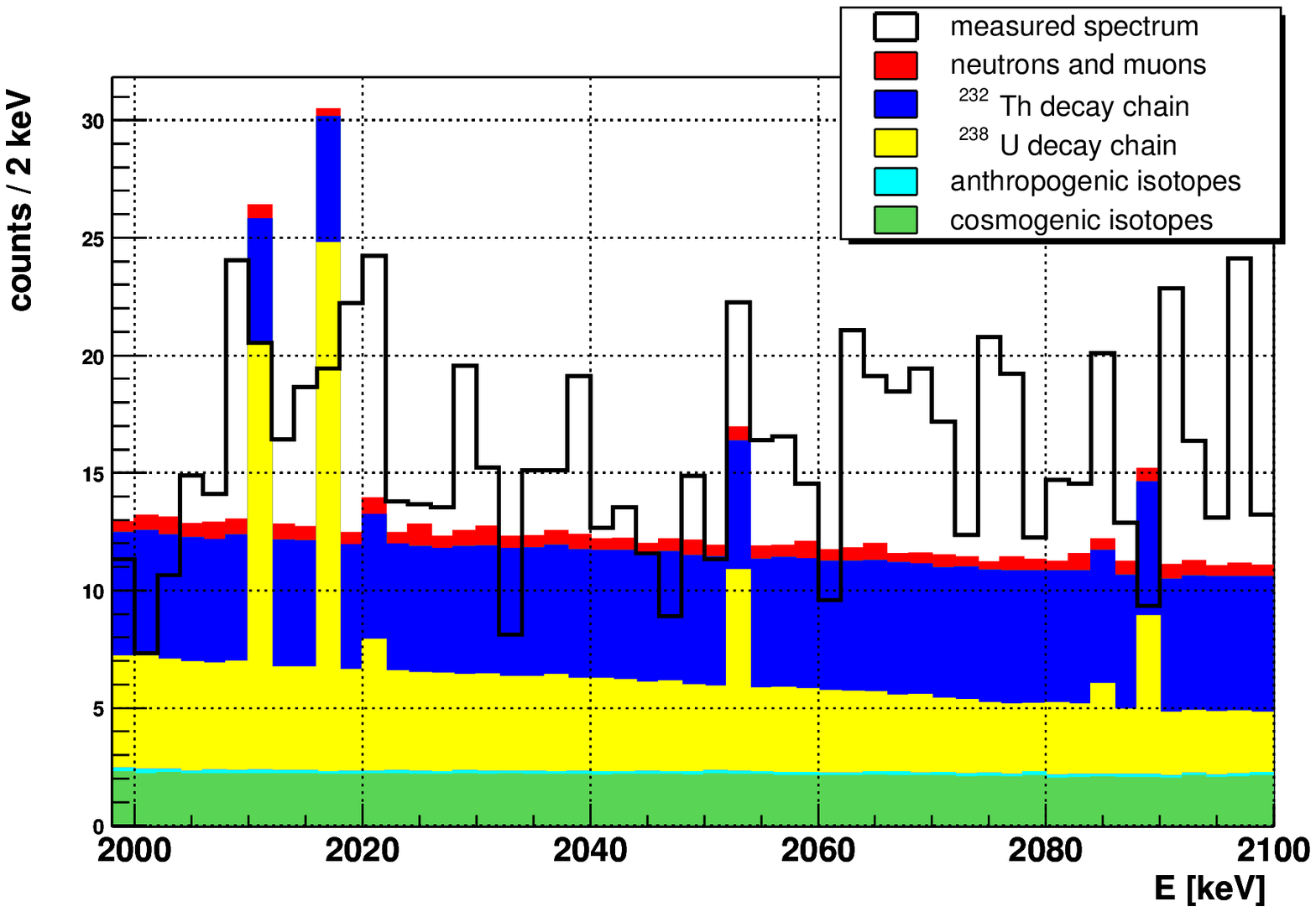}
\caption{\label{picUnder}\rm \small 
	Simulated background of the
	HEIDELBERG-MOSCOW experiment in the energt range from 2000 to 2100 keV 
	with all known background components.
	The black trough-drawn line corresponds to the measured data from
	20.11.1995 to 16.4.2002 (49.59 kg y).}
\end{center}
\end{figure}


	{\it The background around $Q_{\beta\beta}$ from GEANT4 simulation:} 
\hspace{0.2cm}
	Finally the background around $Q_{\beta\beta}$ will be discussed 
	from the side of simulation. A very careful new simulation 
	of the different components of
	radioactive background in the HEIDELBERG-MOSCOW experiment has been
	performed recently by a new Monte Carlo program based on GEANT4 
\cite{cdoerr,ref28}.
	This simulation uses a new event generator for simulation 
	of radioactive
	decays basing on ENSDF-data and describes the decay of arbitrary
	radioactive isotopes including alpha, beta and gamma emission 
	as well as conversion electrons and X-ray emission. 
	Also included in the simulation is the influence of neutrons 
	in the energy range from thermal to high
	energies up to 100 MeV on the measured spectrum. Elastic and inelastic
	reactions, and capture have been taken into account, 
	and the corresponding
	production of radioactive isotopes in the setup. 
	The neutron fluxes and
	energy distributions were taken from published measurements 
	performed in the Gran Sasso. Also simulated was the cosmic 
	muon flux measured in the
	Gran Sasso, on the measured spectrum.
	To give a feeling for the quality of the simulation,
	Fig. 
\ref{picUnderTotal} 
	shows the simulated and the measured spectra for a $^{228}$Th 
	source spectrum for 4 of our five detectors. 
	It should be noted, that this simulation is {\it not} a fit. 
	It is a calculation of the expected spectra on the basis 
	of the knowledge of the geometry of the setup, and its properties 
	of absorption and scattering of gamma radiation, and 
	of the known position and strength of the source.
	The agreement is excellent.

	The simulation of the background of the experiment reproduces  
	a l l observed lines in the energy range between threshold  
	(around 100 keV) and 2020\,keV 
\cite{dietzdiss,cdoerr,ref28}.  
	Fig. 
\ref{picUnder} 
	shows the simulated background in the range
	2000-2100 keV with all  k n o w n  background components. 
	The black solid line corresponds to the measured data 
	in the period 20.11.1995 - 16.4.2002 (55.57\,kg\,y). 
	It should be noted here, that the simulated spectrum 
	is not folded with the energy resolution of the detectors 
	(which explains the sharp 'lines' in the simulated spectrum).

	The background around $Q_{\beta\beta}$ is according 
	to the simulations  f l a t, the only expected lines come 
	from $^{214}$Bi (from the $^{238}$U natural decay chain)
	at 2010.89, 2016.7, 2021.6, 2052.94, 2085.1 and 2089.7 keV. 
	Lines from cosmogenically produced $^{56}$Co (at 2034.76 keV 
	and 2041.16 keV), half-life 77.3 days, are not expected 
	since the first 200 days of measurement of
	each detector are not used in the data analysis. 
	Also the potential contribution from
	decays of $^{77}$Ge, $^{66}$Ga, or $^{228}$Ac, should 
	not lead to signals visible in our
	measured spectrum near the signal at $Q_{\beta\beta}$. 
	For details we refer to  
\cite{ref28}.

	The structures around 2028 keV, 2066 keV and 2075 keV seen - 
	as also the
	$^{214}$Bi lines -  in practically all Ge experiments (see above),
	cannot be identified at present. 

	{\it Conclusion:} \hspace{0.2cm}
	Concluding, additional support has been given for the evidence 
	of a signal for neutrinoless double beta decay, by
	showing consistency of the result - for the signal,  a n d  for the
	background - with other double beta decay experiments using
	non-enriched or enriched Germanium detectors. In particular it has been
	shown that the lines seen in the vicinity of the signal (including
	those which at present cannot be attributed) are seen also in
	the other experiments. This is important for the correct 
	treatment of the background. 
	Furthermore, the sensitivity of the peak identification
	procedures has been demonstrated by extensive statistical simulations.
	It has been further shown by new extensive simulations of the expected
	background by GEANT4, that the background around 
	$Q_{\beta\beta}$ should be flat, and
	that no known gamma line is expected at the energy of $Q_{\beta\beta}$.
	The 2039 keV signal is seen  \mbox{o n l y}  in
	the HEIDELBERG-MOSCOW  experiment, which has a factor of 
	at least 10, but in general 
	\mbox{m u c h} more, statistics than all other double beta experiments.



{\small
\begin{thebibliography}{0}

\bibitem{evid1} 
        H.V. Klapdor-Kleingrothaus et al.
        hep-ph/0201231 and {\it Mod. Phys. Lett.} 
        {\bf A 16} (2001) 2409-2420.

\bibitem{evid2}
        H.V. Klapdor-Kleingrothaus, 
        A. Dietz and I.V. Krivosheina,
        Part. and Nucl. {\bf 110} (2002) 57-79.

\bibitem{evid3} 
        H.V. Klapdor-Kleingrothaus, A. Dietz and I.V. Krivosheina,  
        {\it Foundations of Physics} {\bf 31} (2002) 1181-1223 
        and Corrigenda, 2003 home-page: \\
        http://www.mpi-hd.mpg.de/non\_acc/main\_results.html

\bibitem{evid4}
        H.V. Klapdor-Kleingrothaus,
        hep-ph/0303217. 

\bibitem{KK-OULU02}
        H.V. Klapdor-Kleingrothaus, in Proc. of BEYOND'02, 
	Third International Conference on Physics Beyond the Standard Model, 
	Oulu, Finland, June 2 - 7, 2002, Springer, 2003, 
	ed. H.V. Klapdor-Kleingrothaus. 

\bibitem{60years}
        H.V. Klapdor-Kleingrothaus, 
                {\sf "60 Years of Double Beta Decay - From
        Nuclear Physics to Beyond the Standard Model"}, 
        {\it World Scientific, Singapore} (2001) 1281~p.

\bibitem{con1}
        H.V. Klapdor-Kleingrothaus, H. P\"as and A.Yu. Smirnov, 
        {\it Phys. Rev.} {\bf D 63} (2001) 073005 and 
        {\it hep-ph/}{\bf 0003219}.       

\bibitem{con2}
        H.V. Klapdor-Kleingrothaus, U. Sarkar, Mod. Phys. Lett. 
        {\bf A 16} (2001) 2460-2482.

\bibitem{con3}
        H.V. Klapdor-Kleingrothaus, U. Sarkar,
        hep-ph/0304032.
                      
\bibitem{cmb1}
        D.N. Spergel \etal, astro-ph/0302209.

\bibitem{cmb2}
        S. Hannestad, astro-ph/0303076.

\bibitem{cmb3}
        A. Pierce, H. Murayama, astro-ph/0302131.

\bibitem{cray}
        Z. Fodor \etal, JHEP (2002) 0206046 or 
        hep-ph/\-0203198 and hep-ph/0210123;
        D. Fargion \etal, Proc. DARK 2000, Heidelberg, 
        ed.\\
	 H.V. Klapdor-Kleingrothaus (Springer, Heidelberg, 2001)
        455 and Proc.\\ BEYOND02, Oulu Finland,
        ed. H.V. Klapdor-Kleingrothaus
        (IOP Bristol, 2003); H. P\"as and H. Weiler, 
        Phys. Rev. {\bf D 63} (2001) 113015.

\bibitem{tritium}
        J. Bonn \etal, Nucl. Phys. {\bf B (Proc. Suppl.) 91} (2001) 273.

\bibitem{ma}
        K.S. Babu, E. Ma, J.W.F. Valle, hep-ph/0206292.

\bibitem{mohap}
        R. Mohapatra, M. K. Parida, G. Rajasekaran, hep-ph/0301234.

\bibitem{qvalue}
        G. Douysset \etal,
        {\it Phys. Rev. Lett.} {\bf 86} (2001) 4259 - 4262.

\bibitem{toi}
        R.B. Firestone and V.S. Shirley, 
        Table of Isotopes, 8-th Edition, 
        {\it John Wiley and Sons}, Incorp., N.Y. (1998).

\bibitem{oleg}
        H.V. Klapdor-Kleingrothaus \etal, in press, NIM 2003.

\bibitem{kk0205}
        H.V. Klapdor-Kleingrothaus, hep-ph/0205228.

\bibitem{gamma}
        G. Gilmore, J. Hemingway, 
        ``Practical Gamma-Ray Spectrometry'', 
        Wiley and Sons (1995).

\bibitem{caldwell}
        D. Caldwell \etal, J. Phys. {\bf G 17} (1991) S137-S144.

\bibitem{vasenko}
        A.A. Vasenko \etal, Mod. Phys. Lett. {\bf A 5} (1990) 1299,
        and I. Kirpichnikov, Preprint ITEP (1991).

\bibitem{igex}
        C.E. Aalseth \etal, Yad. Fiz. {\bf 63} (2000) 129. 

\bibitem{Dum-Comm01}
	A. Aalseth et.al, hep-ex/0202018 (vers. 1) and Mod. Phys. Lett. 
	{\bf A 17} (2002) 1475-1478 and 
	F. Feruglio et.al., Nucl. Phys. {\bf B 637} (2002), 345, and 
	Yu. G. Zdesenko et. al., Phys. Lett. {\bf B 546} (2002) 206.

\bibitem{igex2}
        C.E. Aalseth \etal, Phys. Rev. {\bf D 65} (2002) 092007.


\bibitem{dietzdiss}
        A. Dietz, Dissertation, University of Heidelberg, 2003.

\bibitem{random}
        CERN number generators (see e.g. \\
	{\tt http://root.cern.ch/root/html/TRandom.html})

\bibitem{kirch}
        I. Kirpichnikov, talk at Conf. on Nucl. Phys Russ. Acad. Sci.,
        Moscow, Dec. 2 (2002), and private communication , Dec. 2002.

\bibitem{cdoerr}
        C. D\"orr, diploma thesis, University of Heidelberg, 2002, 
	unpublished.
 
\bibitem{ref28}
        H.V. Klapdor-Kleingrothaus \etal, to be published.

\end{thebibliography}
}
\end{document}